\documentclass[prb,twocolumn,preprintnumbers,superscriptaddress]{revtex4}

\usepackage{graphicx,epstopdf}% Include figure files
\usepackage{amssymb,amsmath}
\usepackage{dcolumn}% Align table columns on decimal point
\usepackage{bm}% bold math
\usepackage{color}
\usepackage{enumerate}
\usepackage[colorlinks=true,citecolor=blue]{hyperref}
\hypersetup{colorlinks=true,citecolor=blue,linkcolor=blue,urlcolor=blue}

\usepackage{soul}

\begin{document}

\title{Tunable strongly interacting dipolar excitons in hybrid perovskites}

\author{D.A. Baghdasaryan}
\affiliation{Russian-Armenian University, Yerevan 0051, Armenia}

\author{E.S. Hakobyan}
\affiliation{Russian-Armenian University, Yerevan 0051, Armenia}

\author{D.B. Hayrapetyan}
\affiliation{Russian-Armenian University, Yerevan 0051, Armenia}

\author{I.V. Iorsh}
\affiliation{ITMO University, St. Petersburg 197101, Russia}

\author{I. A. Shelykh}
\affiliation{Science Institute, University of Iceland IS-107, Reykjavik, Iceland}
\affiliation{ITMO University, St. Petersburg 197101, Russia}

\author{V. Shahnazaryan}
\affiliation{ITMO University, St. Petersburg 197101, Russia}

\begin{abstract}
We study theoretically the excitonic nonlinearity in hybrid organic-inorganic Ruddlesden–Popper perovskite thin films. 
The composite layered structure of these materials allows for flexible modulation of their excitonic response between the limiting cases of single atomic layer and wide quasi- two-dimensional quantum well.
In particular, we demonstrate that transverse electric field leads to the spatial separation of charge carriers within the inorganic layer, giving rise to strongly interacting excitons possessing built-in dipole moment. Combined with exciton binding energy of the order of hundreds of meVs, this makes hybrid perovskites  an optimal platform for tailoring of nonlinear optical response at reduced dimensionality.

\end{abstract}
%\date{\today}

\maketitle

\medskip

%\textbf{Keywords:} {Ruddlesden–Popper perovskites (RPPs), exciton energy, excitonic sates, electric field.}

%\medskip

\section{Introduction}

The properties of excitons in semiconductors are essentially modified in the case of the spatial separation between an electron and a hole \cite{Lozovik1976}.
The corresponding state, referred to as indirect or dipolar exciton, is characterized by an increased lifetime due to the reduced overlap of electron and hole wavefunctions \cite{Butov2017}. 
Another consequence of the spatial separation of charge carriers is nonzero built-in electric dipole moment of excitons, which results in the long range exciton-exciton dipolar interactions \cite{Laikhtman2009,Kyriienko2012}.
Their presence manifests itself in a variety of quantum collective effects, including excitonic Bose-Einstein condensation \cite{Butov2002,High2012}, superfluidity \cite{Anakine2017}, formation of exotic dipolar liquid phases \cite{Misra2018,Hubert2019}, and qualitative modification of transport phenomena \cite{Ivanov2002,Winbow2011,Cohen2011,Fedichkin2015,Dorow2016,Shahnazaryan2021,Chiaruttini2021}.
Formation of robust dipolar excitons was reported in a variety of experimantal geometries, which include double quantum wells based on GaAs \cite{Butov1999} or GaN \cite{Fedichkin2015}, and bilayers of atomically thin transition metal dichalcogenides \cite{Fogler2014,Calman2018}.
The presence of dipolar excitons can substantially modify an optical response of a system, and in the regime of strong light-matter coupling can lead to the emergence of the so called dipolaritons \cite{Cristofolini2012,Rosenberg2018,Togan2018}. 

It was recently proposed, that layered two-dimensional (2D) Ruddlesden-Popper  organic–inorganic metal halide perovskites (RPP) \cite{Mitzi1994} can represent a  promising platform for excitonics and polaritonics. These materials are characterized by the chemical formula  ${A_2}{A'_{n - 1}}{M_n}{X_{3n + 1}}$, where $n$ is the number of perovskite layers, related to a thickness of a quantum well,  $M$ denotes a metal, $A$ and $A'$ are cations, and $X$ is a halide.
The current state of the fabrication techniques allows to vary the parameter $n$ from $n=1$ to $n\rightarrow\infty$, which makes possible a controllable crossover between atomically thin and bulk limits.
The optical response of thin RPP films demonstrates the presence of sharp exciton peaks even at room temperatures  \cite{Ahmad2015,Straus2018,Li2019,Marongiu2019,Deng2020}, 
with corresponding exciton binding energy up to 500 meV \cite{Tanaka2005,Yaffe2015,Mauck2019}.
The variation of the number of inorganic layers essentially modifies the excitonic states due to the modulation of Coulomb interaction,  effects of bandgap renormalization, and change in the exciton-phonon coupling \cite{Saparov2016,Gong2018,Straus2018,Blancon2018,Quan2019}.
In addition, the nonlinear optical response, measured as blueshift of an exciton \cite{Huang2017,Abdelwahab2019,Ohara2019}, was reported to be substantially higher, then in conventional excitonic materials.

%In particular, we will consider structures of the form ${({BA})_2}{( {MA})_{n - 1}}P{b_n}{I_{3n + 1}}$, where $BA$ and $MA$ stand for butylammonium and methylammonium respectively.

Here we demonstrate the application of an external electric field in  RPP materials can result in further enhancement of excitonic nonlinear response. 
Indeed, electric field essentially modifies the internal structure of exciton states,  inducing a built-in dipole moment, which changes the character of exciton-exciton interactions from short range dominated by electronic and hole exchange, to long range dominated by dipole-dipole repulsion.

The paper is organized as follows. 
In Sec.~\ref{sec:exciton} we provide the description of excitons in considered structure and analyze the impact of transverse electric field on wavefunctions of exciton states. 
In Sec.~\ref{sec:XXint} we present the results of the calculations of exciton-exciton scatterings in the presence of an electric field, demonstrating a substantial increase of optical nonlinearity.
Sec.~\ref{sec:conc} summarizes the obtained results.

\section{Excitonic states in RPP layer}\label{sec:exciton}
\begin{figure}
    \includegraphics[width=1\linewidth]{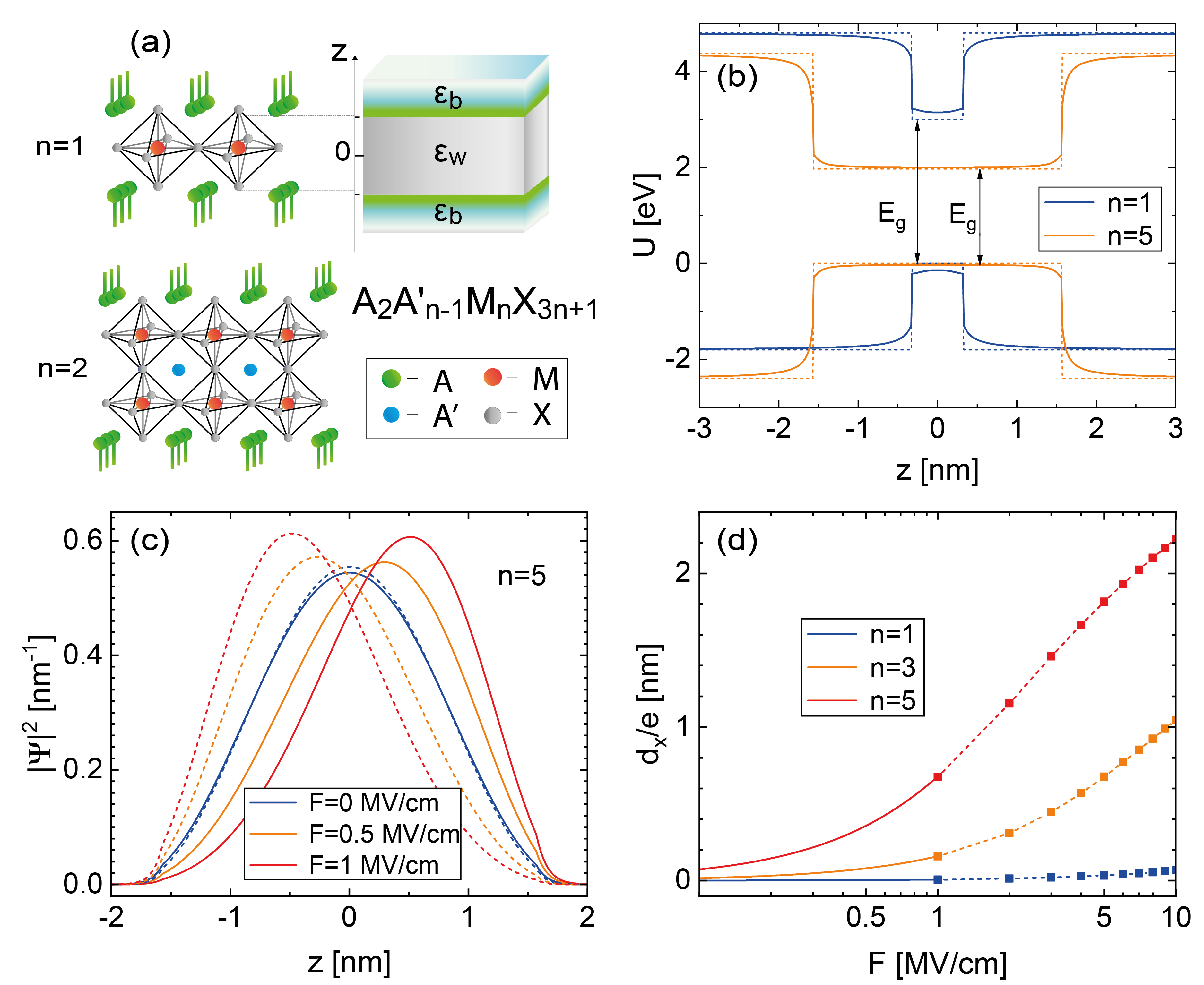}
    \caption{(a) The sketch of RPP structure containing one (top) and two (bottom) inorganic layers. 
    (b) The profile of confinement potential along the stacking direction. The thin dashed lines correspond to the model of square well, and the solid lines illustrate the total confinement accounting for the self-induced polarization effect. 
    (c) The electron (solid lines) and hole (dashed line) wave functions along the stacking direction for different values of the electric field. The number of inorganic layers is taken as $n=5$. 
    (d) The exciton dipole moment versus the strength of the electric field for different values of $n$. 
    }
    \label{fig:psi_z}
\end{figure}
\begin{figure}
    \includegraphics[width=1\linewidth]{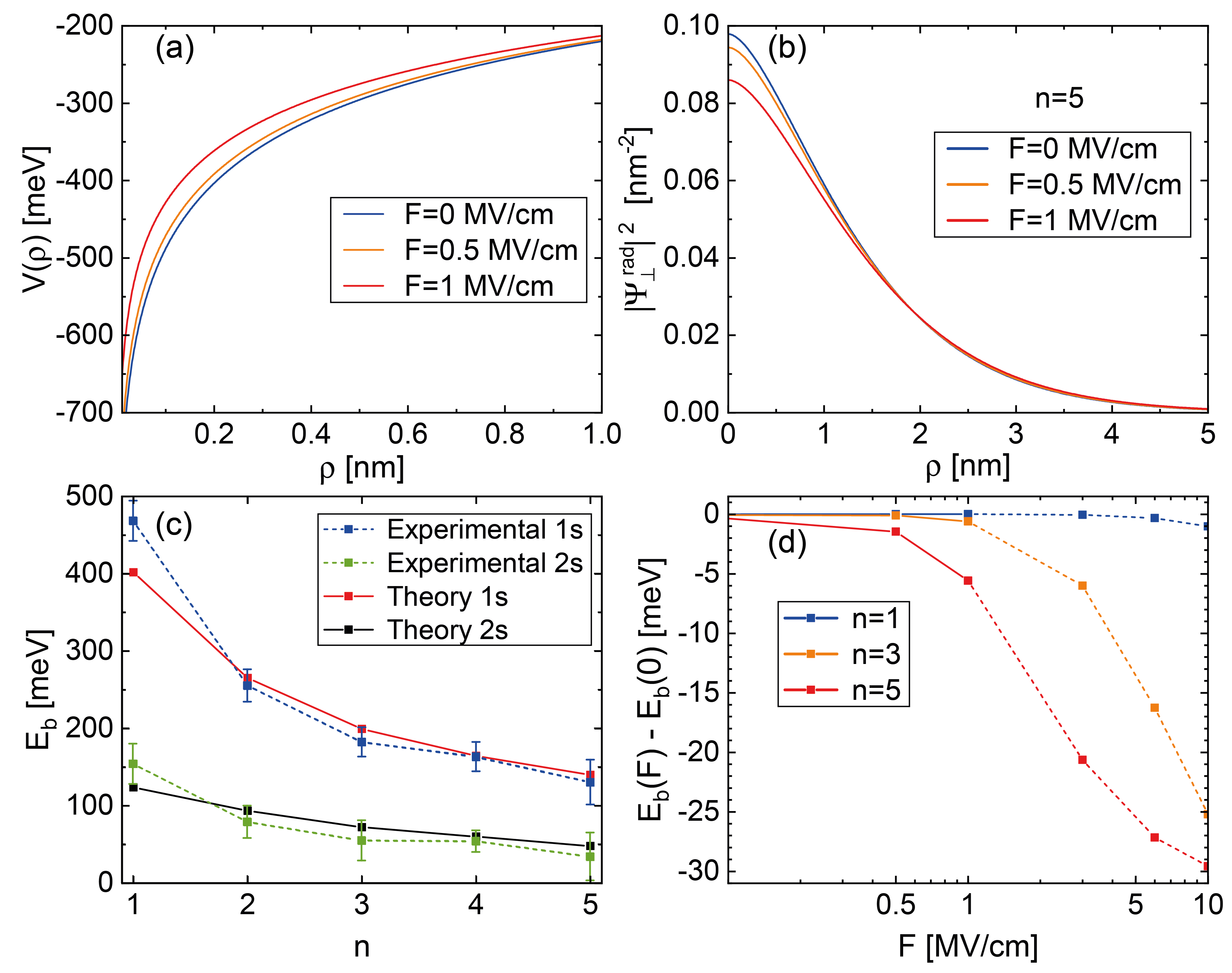}
    \caption{
    (a) The radial dependence of an effective in-plane Coulomb interaction $V(\rho)$ for different values of the electric field.
    (b) The radial probability distribution of 1s state 
    %wave function 
    for different values of electric field for $n=5$. 
    (c) The energy of $1s$ and $2s$ exciton states versus the number of the layers. The dashed curves corresponds to the experimental data of the Ref. \cite{Blancon2018}.
    (d) The reduction of exciton binding energy versus the electric field for different numbers of the layers.
    For a narrow well the binding energy demonstrate only a tiny change due to the tight confinement of an electron and a hole.
    }
    \label{fig:Xrho}
\end{figure}
\subsection{The model}

We consider an exciton state in RPP layer in the presence of an external electric field along the stacking direction. 
The corresponding structure with one and two inorganic perovskite layers is schematically shown in Fig.~\eqref{fig:psi_z} (a).
The thickness of an organic spacer is assumed to be large enough to neglect the impact of any kind of the superlattice effects. 
The Hamiltonian of an interacting electron-hole pair reads:
\begin{align}
    \label{eq:Ham0}
    \hat{H} =   - \frac{\hbar^2}{2\mu _e}\Delta_{{\bm r} e} 
    - \frac{\hbar^2}{2\mu _h}\Delta_{ {\bm r}h} 
    + U\left( z_e,z_h,{\bm \rho} \right) +e F (z_h - z_e),
\end{align}
where $\mu_{e[h]}$ is an electron [hole] effective mass, ${\bm r}_i = \left( {\bm \rho}_i, z_i \right)$, $i = e,h$, 
${\bm \rho} = {\bm \rho}_e - {\bm \rho}_h$ is the in-plane relative coordinate,  
$e$ is the elementary charge 
and $F$ is the electric field strength.

The potential energy term can be written as:
\begin{align}
    \label{eq:U_rho}
    U\left( z_e,z_h, {\bm \rho}  \right) =
    U_e(z_e) + U_h(z_h) 
    + V\left( z_e ,z_h,{\bm \rho}  \right),
\end{align}
where $ V\left( z_e ,z_h,{\bm \rho}  \right)$ is the Coulomb interaction between an electron and a hole. ${U_i(z_i)}$ are one-particle confining potential energies appearing due to the conduction and valence band offsets, together with the self-energy correction provided by the mismatch of the dielectric constants in the regions of a barrier and a well: 
\begin{equation}
    \label{U_z}
    U_i\left( z_i \right) = U_{\rm conf}^i (z_i) + U_{\rm self}^i (z_i).
\end{equation}
Here ${U_{\rm self}^i}$
is the self-energy potential, resulting from the interaction of carriers with their image charges in a dielectric medium.
${U_{\rm conf}^i}$ is the confining potential defined by the band offset between the perovskite and its surrounding material, approximated by square well:
\begin{align}
    \label{eq:U_conf}
    U_{\rm conf}^i (z_i) =
    \begin{cases}
        U_0^i, \quad  z_i < -l/2, \quad z_i < l/2 ,  \\
        0, \quad -l/2 \leq z_i \leq l/2.
    \end{cases}
\end{align}
$l=n l_0$ denotes the width of a perovskite inorganic layer, i. e. the width of  quantum well; and $l_0= 0.6$ nm is the width of single atomic layer.

\subsection{Polarization-induced confinement potential}

The correction to the confinement potential arising from the self-induced polarization can be accounted for within the image charge method \cite{PanofskyBook}, which was previously applied for the description of the polarization effect in quantum wells \cite{Kumagai1989}. The emergent image charges are located at 
\begin{equation}
    (-1)^m z + m l, \qquad m =  0, \pm 1, \pm 2 \ldots ,
\end{equation}
where $z$ denotes the position of the real charge. 
The values of corresponding image charges are determined via the continuity conditions for the electrostatic potential and the normal component of the displacement vector at the interfaces, and have a form
\begin{equation}
    \begin{cases}
    e_m = \kappa^m, \quad \kappa = \frac{\epsilon_w-\epsilon_b}{\epsilon_w+\epsilon_b}, \quad z <l/2, \\
    e^\prime_m =  \frac{2\epsilon_b}{\epsilon_w+\epsilon_b}{e}_{m}, \quad  z \geq l/2,
    \end{cases}
    \quad m =  0, \pm 1, \pm 2 \ldots .
\end{equation}
Here $\epsilon_w = 9$ is the dielectric constant of inorganic layer, $\epsilon_b=2.2$ is dielectric constant of inorganic spacer \cite{Blancon2018}. 

The self-induced confinement potential is calculated as
\begin{align}
    U_{\rm self}^i &= 
    \sum_{m=\pm1,\pm2, ...}
    \frac{\kappa^{|m|} e^2 }{2 \epsilon_w  |z - {\left( { - 1} \right)^m}{z } + ml|},
    \quad |z| < l/2,
     \notag \\
    \begin{split}
    U_{\rm self}^i & = \frac{2 \epsilon_w}{\epsilon_w + \epsilon_b} \sum_{m=0}^{\infty} \frac{\kappa^{2m+1} e^2}{(\epsilon_w + \epsilon_b)
     |2z + (2m+1)l |} \\
    & - \frac{\kappa e^2}{2 \epsilon_b | 2 z - l|},
    \quad z > l/2 ,
    \end{split} 
     \notag \\
    U_{\rm self}^i(z) & = U_{\rm self}^i(-z),
    \quad z < -l/2 .
\end{align}
The above expressions have divergencies at the layer interfaces, which can be lifted by application of the so called "shifted mirror faces" procedure \cite{Lang1973,Kumagai1989}. 
The values of confinement potentials for different number of the layers are taken from Ref. \cite{Blancon2018}. 
The corresponding energy profiles for representative cases $n=1$ and $n=5$ are presented in the Fig.~\ref{fig:psi_z} (b).

\subsection{Ansatz for an excitonic wavefunction}

We apply the well-established procedure of the separation of slow in-plane dynamics and fast dynamics in the growth direction, which was previously successfully employed to describe excitonic states in perovskite superlattices \cite{Muljarov1995}. 
We use the following ansatz for excitonic wavefunction:
\begin{equation}
    \label{eq:xciton_wf}
    \Psi_{\bf Q} \left( {\bm r}_e,{\bm r}_h \right) = \psi^e(z_e) \psi^h (z_h) \psi_{\perp} ({\bm \rho}) \Phi_{\bm Q}({\bm R}),
\end{equation}
where $\psi^i(z_i)$ are electron and hole wave functions in $z$ direction, and $\psi_{\perp}$ is the wave function of their in-plane relative dynamics.  Here ${\bm R} = \beta_e {\bm \rho_e} + \beta_h {\bm \rho_h}$ with $\beta_{e[h]} = \mu_{e[h]}/(\mu_e+\mu_h)$, and
\begin{equation}
    \Phi_{\bm Q}\left( {\bm R} \right) = \frac{1}{\sqrt{A}} e^{i {\bm Q} {\bm R} }
\end{equation}
is the wave function corresponding to the in-plane motion of a center of mass, characterized by wave vector ${\bm Q}$. Here $A$ is the normalization area. 

The wavefunctions $\psi^i(z_i)$ are defined by the following equations:
\begin{align}
    \label{eq:Schro_z}
    \left( - \frac{\hbar^2}{2\mu_i} \frac{{\rm d}^2}{{\rm d} z_i^2} 
    + U_i (z_i) \pm eF z_i \right) \psi ^i ( z_i)
    = E_z^i \psi^i( z_i).
\end{align}
This approximation is valid, if characteristic energy of the dimensional quantization of individual electrons and holes is larger then characteristic exciton binding energy, 
$\pi^2 \hbar^2 / (2 m_i (nl_0)^2) \gg e^2 / (4\pi\varepsilon_0\varepsilon_w n l_0) $, which in our case holds for $n\leq 6$. 

Fig.~\ref{fig:psi_z} (c) shows  wave functions of the charge carriers in $z$-direction for different values of external electric field, which drags positive and negative charge carriers in opposite directions. Consequently, an exciton acquires a built-in exciton dipole moment along the $z$ direction, calculated as:
\begin{equation}
    d_X = e \int z \left( \left| \psi^e (z) \right|^2 - \left| \psi^h(z) \right|^2 \right)  dz.
\end{equation}
The electric field dependence of an exciton dipole moment is illustrated in Fig.~\ref{fig:psi_z} (d).
Evidently, for the monolayer case ($n=1$) due to the strong confinement in z direction the impact of an electric field is negligible.
For wider quantum wells at the limit of realistic field values $F = 1$ MV/cm one can see a significant separation of charges of about $0.7$ nm, which corresponds to a dipole moment of about 35 D.

\subsection{Exciton in-plane wave function}

The in-plane relative motion of an exciton is described by the following equation:
\begin{align}
    \label{Schr_r}
    - \frac{\hbar ^2}{2\mu} \left( \frac{{\rm d}^2}{ {\rm d} \rho^2} + \frac{1}{\rho} \frac{{\rm d}}{{\rm d} \rho} - \frac{m^2}{\rho^2} \right) \psi_{\perp}^{\rm rad} + \left( E_b + V_{\rho}(\rho) \right) \psi_{\perp}^{\rm rad} = 0 ,
\end{align}
where $\mu=\mu_e \mu_h/(\mu_e+\mu_h)$ is the reduced mass of an exciton , $\psi_{\perp} (\rho,\varphi) = \psi_{\perp}^{\rm rad} (\rho) e^{im\varphi}$, $m$ is the magnetic quantum number, and $-E_b$ is the binding energy of an exciton. The effective in-plane potential can be found by averaging in $z$ direction:
\begin{align}
    \label{V_rho}
    V_{\rho} (\rho) =
    \int \left|\psi^e (z_e)\right|^2 \left|\psi^h(z_h)\right|^2
    V\left( z_e,z_h,\rho\right) {\rm d} z_e {\rm d} z_h,
\end{align}
where the electron-hole Coulomb interaction for the particles inside the inorganic layer reads \cite{Kumagai1989}:
\begin{align}
   V\left( z_e,z_h,\rho\right) 
   = - \sum_{m=-\infty}^{\infty} \frac{\kappa^{\mid m\mid} q_e^2}
        {\epsilon_w \sqrt{
            \rho^2 +  ( z_e - \left( - 1 \right)^m z_h + m l)^2}} .
\end{align}
The radial dependence of in-plane Coulomb interaction for different values of perpendicular electric field is depicted in Fig.~\ref{fig:Xrho} (a).
The presence of an electric field leads to the charge separation in $z$ direction,  thus weakening effective in-plane Coulomb attraction. In turn, this results in less bound excitons, as shown in Fig.~\ref{fig:Xrho} (b).
The exciton binding energy in the absence of an electric field is presented Fig.~\ref{fig:Xrho} (c), demonstrating a good agreement with available experimental data \cite{Blancon2018}. Here we consider $s$ states only ($m=0$).

Electric field induced reduction of the exciton binding energy is shown in Fig.~\ref{fig:Xrho} (d). We note that considerable effect appears only for the large number of layers, necessary for the effective field-induced charge separation.

%%%%%%%%%%%%%%%%%%%%%%%%%%%%%%

\section{Exciton-exciton interaction}\label{sec:XXint}

\begin{figure}
    \includegraphics[width=1.\linewidth]{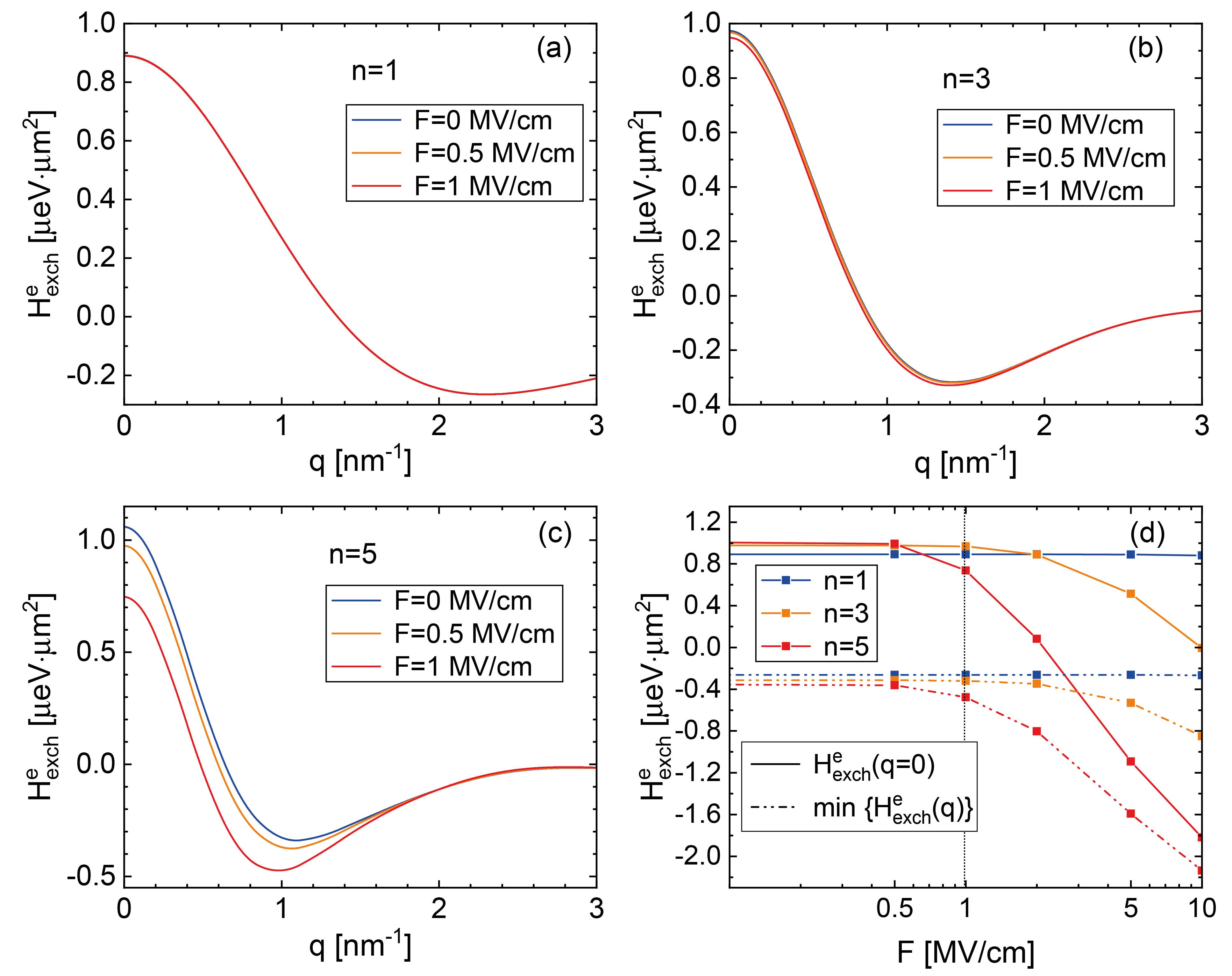}
    \caption{(a), (b), (c) The dependence of the exciton-exciton exchange interaction on transferred momentum for different values of an electric field.  The number of layers is $n=1,3,5$ for panels (a), (b), (c) respectively. 
    (d): The dependence of the exchange interaction on electric field for different number of layers. 
    %At the region $F>1$ MV/cm 
    The solid curves correspond to $q=0$, and the dashed ones to $q$ at which the scattering matrix element reaches its minimum. Note, that the value of a scattering matrix element at $q=0$ can, in principle change sign, as it can be seen from the solid red curve corresponding to $n=5$. However, the values of the electric field, necessary for that, are beyond experimentally accessible.
    }
    \label{fig:XXexch}
\end{figure}
\begin{figure}
    \includegraphics[width=1.\linewidth]{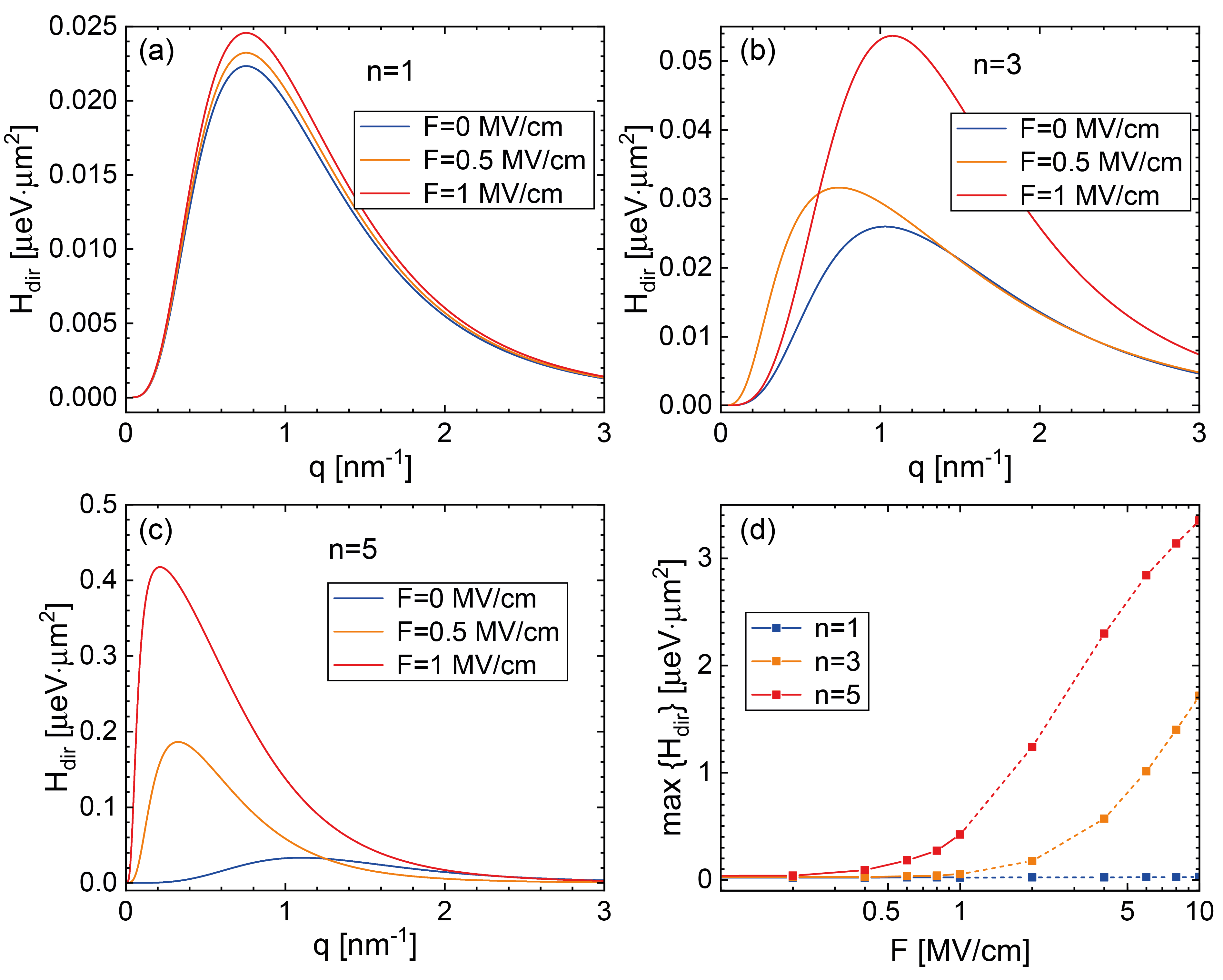}
    \caption{(a), (b), (c): The dependence of the direct exciton-exciton interaction on exchange momentum for different values of an electric field.  The number of the layers is $n=1,3,5$ in the panels (a), (b), (c) respectively.  
    (d): The dependence of the maximum of the direct interaction on electric field for different number of the layers. 
    }
    \label{fig:XXdir}
\end{figure}
\begin{figure}
    \includegraphics[width=1.\linewidth]{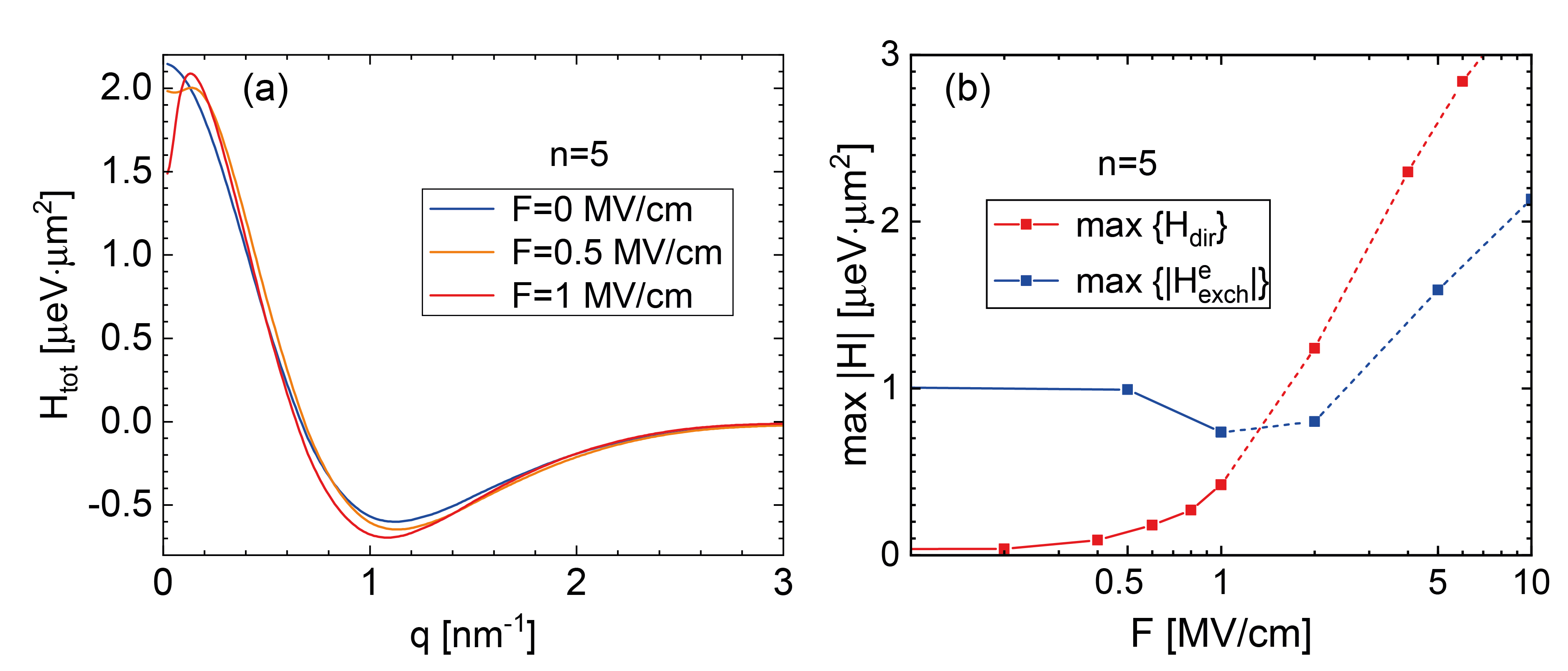}
    \caption{ (a) The  dependence of the total exciton-exciton interaction on transferred momentum. (b) The corresponding dependence  of the maxima of direct and exchange interactions on electric field.
    }
    \label{fig:XXtot}
\end{figure}
 \begin{figure}
     \includegraphics[width=1.\linewidth]{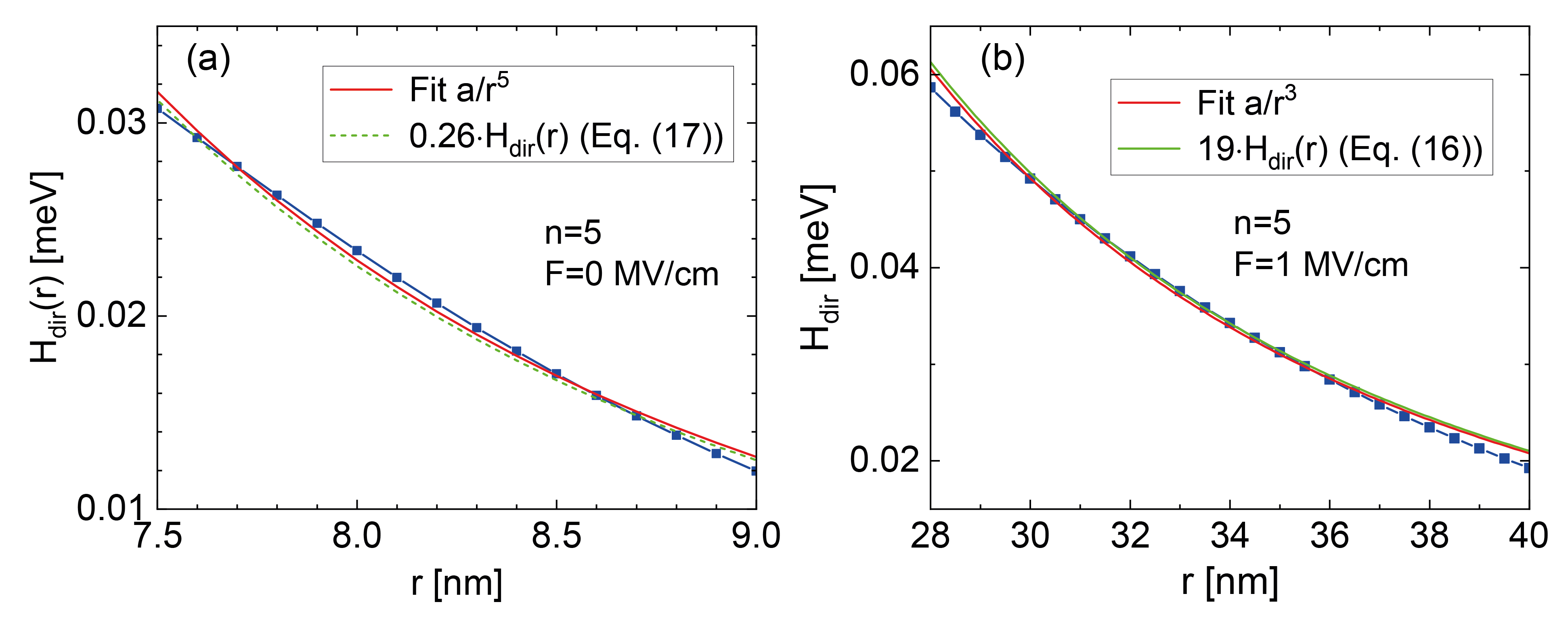}
     \caption{ The real space dependence of the direct exciton-exciton interaction. (a) The interaction in the absence of an electric field, demonstrating an $r^{-5}$ scaling. (b) The interaction of electrically polarized dipolar excitons, demonstrating $r^{-3}$ scaling. 
     }
     \label{fig:Vr}
 \end{figure}

The wavefunctions of the excitonic states obtained in the previous section can be used for the calculation of the exciton-exciton scattering elements in Born approximation \cite{Ciuti1998}. Restricting the treatment to case of parallel spin alignment, the interaction process with transfer of a wave vector $\mathbf{q}$ can be schematically represented as
\begin{equation}
    (\mathbf{Q}) + (\mathbf{Q}) \rightarrow (\mathbf{Q}+\mathbf{q})+(\mathbf{Q}'-\mathbf{q}).
    \label{scattering}
\end{equation}
The corresponding scattering amplitude is the sum of the four interaction channels, namely direct interaction and exciton, electron and hole exchanges:
\begin{eqnarray}
    H( {\bm \Delta Q}, {\bm q})= &H_{{\rm dir}} ( {\bm q}) 
    +H^X_{{\rm exch}} ({\bm \Delta Q}, {\bm q}) \notag\\ 
    +&H^e_{{\rm exch}} ({\bm \Delta Q}, {\bm q})
    +H^h_{{\rm exch}} ({\bm \Delta Q}, {\bm q}),
\end{eqnarray}
where ${\bm \Delta Q} = {\bm Q} - {\bm Q}^\prime$. 
The explicit expressions for matrix elements are presented in Appendix A.

It was shown previously \cite{Ciuti1998,Tassone1999,Shahnazaryan2016, Shahnazaryan2017} that in the absense of an external electric field in a wide region of the exchanged wave vectors $q\leq 1/a_B$, where $a_B = \langle \psi_{\perp}|\rho|\psi_{\perp}\rangle$ is the exciton Bohr radius, exchange contribution dominates, and the direct term can be safely neglected. 
The direct interaction  becomes dominant for large values of $\textbf{q}$, governing the long range behavior of the interaction. 
On the other hand, the presence of dipole momentum can result in the domination of the direct interaction even at small exchange momenta \cite{Kyriienko2012}.
Here we found that the electric field-induced dipole moment essentially modifies the character of both direct and exchange exciton-exciton interactions.

We calculate the exchange interaction matrix element exploiting  the  multi-dimensional  Monte-Carlo  integration  \cite{Hahn2005}.   
The results are shown in Fig.~\ref{fig:XXexch}. For a narrow quantum well an electric field has negligible impact on the exchange interaction, as it is shown in Fig.~\ref{fig:XXexch} (a). 
The increase of the number of the layers in the absence of an electric field results in a minor enhancement of the exchange interaction. We attribute this to the increase of the Bohr radius and, consequently, the interaction cross-section.
For a wider RPP layer with $n=5$  the increase of an electric field leads to a moderate decay of the interaction strength (Fig.~\ref{fig:XXexch} (b),(c)). On the other hand, the attraction region at intermediate momenta $q>0$ becomes larger and comparable with repulsive maximum at $q=0$.
In Fig.~\ref{fig:XXexch} (d) we present the electric field dependence of the exchange interaction, which clearly indicates the domination of the attraction trend at large values of an electric field, which are beyond the experimental reachability ($F> 1$ MV/cm).
We mention that such attraction was earlier reported for a system of coupled quantum wells \cite{Kyriienko2012}, and is analogous to the exchange interaction between excitons in the excited states in conventional quantum wells \cite{Shahnazaryan2016} and atomically thin transition metal dichalcogenides \cite{Shahnazaryan2017}.

The results of the calculation for the direct interaction are shown in Fig.~\ref{fig:XXdir}. 
While for the case of a narrow perovskite layer the presence of an electric field has a minor impact, for the wider layers it strongly enhances the direct interaction (cf. Fig.~\ref{fig:XXdir} (a)-(c)). 
Such behavior is explained by larger exciton dipole moment in wide layers in the presence of an electric field (see Fig.~\ref{fig:psi_z} (d)).
Despite the strong modulation of the interaction strength, the electric field preserves the dome-shaped character of the curve, describing the dependence on transferred  momentum. 
Fig.~\ref{fig:XXdir} (d) illustrates the  dependence of the direct interaction on external electric field or different number of inorganic layers. 
The flattening of the interaction enhancement slope at large electric fields of about $F \sim 10$ MV/cm is attributed to the saturation of  exciton dipole momentum due to the confinement in $z-$ direction.

Fig.~\ref{fig:XXtot} (a) illustrates the dependence of the total interaction on transferred momentum for a wide perovskite layer with $n=5$ in the presence of an electric field.
The corresponding dependencies of the maxima of the exchange and direct interactions on  electric field strength are shown in Fig.~\ref{fig:XXtot} (b),  clearly demonstrating a crossover from the exchange dominated regime to one dominated by the direct interaction with increase of an electric field. 

In order to understand further a qualitative impact of an electric field on the exciton-exciton direct interaction, we shift to the real space domain.
While exciton-exciton exchange interaction is of short-range nature, the direct interaction clearly demonstrates a long-range behavior.
At the distances $r \gg a_B$ the multipole expansion of direct interaction for conventional quantum wells yields \cite{Schindler2008}
\begin{align}
    \label{eq:Vdirr}
    H_{\rm dir}(r) = \frac{1}{4\pi\varepsilon_0\varepsilon_w} 
    \left[ \frac{d_X^2}{r^3} \right. \notag \\
    \left. + \frac{e^2}{r^5} \left( \frac{9}{4} \langle (r\cos\varphi)^2 - z^2 \rangle^2 
    +3 \frac{d_X}{e}  \langle z (r\cos\varphi)^2 - z^3 \rangle  \right)\right] ,
\end{align}
where $\langle\rangle$ denotes the averaging over exciton wavefunctions. In the absence of an electric field the wavefunctions are symmetric in $z-$ direction, and the interaction reduces to
\begin{equation}
    \label{eq:Vdirr0}
     H_{\rm dir}(r) \approx \frac{e^2}{4\pi\varepsilon_0\varepsilon_w} 
\frac{81}{64} \frac{a_B^4}{r^5} .
\end{equation}
In Fig.~\ref{fig:Vr} the real space dependence of the direct interaction at large distances is shown. Fig.~\ref{fig:Vr} (a) corresponds to the absence of an electric field.
The calculated data is well fitted by $r^{-5}$ type dependence, and qualitatively agrees  with the estimate of Eq.~\eqref{eq:Vdirr0}. 
In the presence of electric field the interaction becomes of dipolar type, and the leading order of real space dependence is $r^{-3}$. 
The corresponding dependence is shown in Fig.~\ref{fig:XXtot} (b), and is in qualitative agreement with the estimate of Eq.~\eqref{eq:Vdirr}. We attribute the discrepancy in absolute values to essentially modified shape of the Coulomb interaction in RPP system compared with conventional quantum wells.

\section{Conclusion}\label{sec:conc}

We studied the exciton-exciton interaction in hybrid organic-inorganic perovskite thin films. The exciton state is described within the model of quantum well of finite size, where the self-induced polarization effects are fully accounted for both in external confinement potential and electron-hole Coulomb interaction. The transverse electric field leads to a spatial separation of electrons and holes. It was shown that in sufficiently wide inorganic layers this results in the appearance of excitons possessing built-in dipole moment. The presence of a dipole moment essentially enhances the direct interaction between excitons, which can become dominant for sufficiently large values of an external electric field. The predicted enhancement of long range nonlinearity can play substantial role in the emergence of quantum collective phases in the considered system.

\section*{Acknowledgments}

The reported study was funded by RFBR and SC RA, project number 20-52-05005.
IAS acknowledges the support of the Icelandic research fund, project "Hybrid polaritonics" (Project number 163082-051).

%\newpage
\appendix 

\section{Matrix elements for exciton-exciton interaction}

The derivation of interaction matrix elements is analogous to that developed in Ref~\cite{Ciuti1998}, accounting for the quasi-3-dimensional character of excitons.
The symmetrized two exciton wave function reads as
\begin{align}
    \Phi_{\mathbf{Q},\mathbf{Q}'}(\mathbf{r}_e,\mathbf{r}_h,\mathbf{r}_{e'},\mathbf{r}_{h'})=\notag \\
    = \frac{1}{2}\left[ 
    \Psi_{\mathbf{Q}}(\mathbf{r}_e,\mathbf{r}_h) \Psi_{\mathbf{Q}'}(\mathbf{r}_{e'},\mathbf{r}_{h'}) 
    + \Psi_{\mathbf{Q}}(\mathbf{r}_{e'}, \mathbf{r}_{h'})\Psi_{\mathbf{Q}'}(\mathbf{r}_e,\mathbf{r}_h) \right] \notag \\
    - \frac{1}{2}\left[ 
    \Psi_{\mathbf{Q}}(\mathbf{r}_{e'},\mathbf{r}_h)\Psi_{\mathbf{Q}'}(\mathbf{r}_e,\mathbf{r}_{h'})
    + \Psi_{\mathbf{Q}}(\mathbf{r}_e,\mathbf{r}_{h'})\Psi_{\mathbf{Q}'}(\mathbf{r}_{e'},\mathbf{r}_h) \right].
\end{align}
The interaction Hamiltonian averaged over the $z$-direction is
%Using normalization constants of $z$-direction wave function can be written as:
%
\begin{align}
    V_I \left( {\bm \rho}_e, {\bm \rho} _{e^\prime}, {\bm \rho}_h, {\bm \rho}_{h^\prime} \right) = 
    &V_{ee}\left( \left| {\bm \rho}_{e^\prime} - {\bm \rho}_e \right| \right)
    + V_{hh}\left( \left| {\bm \rho}_{h^\prime} - {\bm \rho}_h \right| \right) \notag \\
    - &V_{eh}\left( \left| {\bm \rho}_{h^\prime} - {\bm \rho}_e \right| \right)
    - V_{he}\left( \left| {\bm \rho}_{e^\prime} - {\bm \rho}_h \right| \right),
\end{align}
where
\begin{align}
    \label{v_ee}
    V_{ij}\left( \left| {\bm \rho}_i - {\bm \rho}_j \right| \right) =    
     \int & V_{ij} \left(z_i, z_j, \left| {\bm \rho}_i - {\bm \rho}_j \right|  \right) \notag \\
    & \left| \psi_i^z( z_i) \right|^2 
    \left| \psi_j^z( z_j) \right|^2  {\rm d} z_i {\rm d} z_j .
\end{align}
The matrix element of direct interaction does not depend on initial momenta ${\bm Q}$, ${\bm Q}^\prime$ and is presented as
\begin{align}
    H_{ {\rm dir}}(q)  =  \frac{1}{A} \left[  2 V_q g(\beta_e q) g(\beta_h q) 
    - V^{e}_q g^2(\beta_h q)  
    - V^{h}_q g^2(\beta_e q)  \right],
\end{align}
where
\begin{align}
&g(\tau) =  2\pi \int J_0(\tau \rho)  |\psi_\perp^{\rm rad}(\rho)|^2 \rho {\rm d} \rho , \notag \\
&V_q = 2\pi \int J_0(q \tau)   V_{eh}(\tau)  \tau {\rm d} \tau , \notag\\
&V^{i}_q = 2\pi \int J_0(q \tau)   V_{ii}(\tau)  \tau {\rm d} \tau. 
\end{align}
%
%\begin{align}
%&V_q = 2\pi \int J_0(q \tau)   V(z_e,z_h,\tau) |\psi^{e}(z_e)|^2 |\psi^{h}(z_h)|^2   \tau {\rm d}z_e {\rm d}z_h {\rm d} \tau , \notag\\
%&V^{i}_q = 2\pi \int J_0(q \tau)   V(z_{i},z_{i^\prime},\tau) |\psi^{i}(z_{i})|^2 
%|\psi^{i}(z_{i}^\prime)|^2  \tau {\rm d}z_{i} {\rm d} z_{i^\prime}   {\rm d} \tau .
%\end{align}
%

The exciton exchange term is
\begin{equation}
H^X_{\mathrm{exch}}({\bm \Delta Q}, {\bm q}) =  H_{\mathrm{dir}}({\bm \Delta Q} - {\bm q}).
\end{equation}
The matrix element of electron exchange interaction has a form
\begin{widetext}
\begin{align}
    H_{exch}^e =  - \frac{1}{A}\int & \cos \left[ \beta_e {\bm \Delta Q} \cdot( {\bm x} -{\bm y}_1) + {\bm q}\cdot ( \beta_h {\bm y}_2 - \beta_e {\bm y}_1 -{\bm x}  ) \right]      V_I \left( {\bm \rho}_e, {\bm \rho} _{e^\prime}, {\bm \rho}_h, {\bm \rho}_{h^\prime} \right) \notag \\
    & \psi_\perp^{\rm rad}(x)
    \psi_\perp^{\rm rad}(y_1) \psi_\perp^{\rm rad}(y_2) \psi_\perp^{\rm rad}(|{\bm y}_2 -{\bm y}_1 -{\bm x}|)
    {\rm d}^2 {\bm x} {\rm d}^2 {\bm y}_1 {\rm d}^2 {\bm y}_2 
\end{align}
\end{widetext}
For ${\bf \Delta Q}=0$ one has
$V^X_{{\rm exch}} (0, {\bm q}) = V_{{\rm dir}} ( {\bm q})$.


\begin{thebibliography}{99}

\bibitem{Lozovik1976}
Yu. E. Lozovik and V. I. Yudson, 
%A new mechanism for superconductivity: Pairing between spatially separated electrons and holes,
Sov. Phys. JETP 44, 389 (1976) [Zh. Eksp. Teor. Fiz. 71,
738 (1976)].

\bibitem{Butov2017}
L. V. Butov,
%Excitonic devices, 
\href{https://www.sciencedirect.com/science/article/abs/pii/S0749603616316561}{Superlattices Microstruct. {\bf 108}, 2 (2017)}.

\bibitem{Laikhtman2009}
B. Laikhtman and R. Rapaport
%Exciton correlations in coupled quantum wells and their luminescence blue shift
\href{
https://journals.aps.org/prb/abstract/10.1103/PhysRevB.80.195313}{Phys. Rev. B {\bf 80}, 195313 (2009)}.

\bibitem{Kyriienko2012}
O. Kyriienko, E. B. Magnusson, and I. A. Shelykh,
%Spin dynamics of cold exciton condensates
\href{https://journals.aps.org/prb/abstract/10.1103/PhysRevB.86.115324}{Phys. Rev. B {\bf 86}, 115324 (2012)}.


%IX BEC
\bibitem{Butov2002}
L. V. Butov, C. W. Lai, A. L. Ivanov, A. C. Gossard, and D. S. Chemla,
%Towards Bose–Einstein condensation of excitons in potential traps
\href{https://www.nature.com/articles/417047a}{Nature {\bf 417}, 47–52 (2002)}.

\bibitem{High2012}
A. A. High, J. R. Leonard, A. T. Hammack, M. M. Fogler, L. V. Butov, A. V. Kavokin, K. L. Campman, and A. C. Gossard, 
\href{https://www.nature.com/articles/nature00943}{Nature {\bf 483}, 584–588 (2012)}.

%superfluid
\bibitem{Anakine2017}
R. Anankine, M. Beian, S. Dang, M. Alloing, E. Cambril, K. Merghem, C. Gomez Carbonell, A. Lemaitre, and F. Dubin,
%Quantized Vortices and Four-Component Superfluidity of Semiconductor Excitons
\href{https://journals.aps.org/prl/abstract/10.1103/PhysRevLett.118.127402}{Phys. Rev. Lett. {\bf 118}, 127402 (2017)}.

%dipolar liquid
\bibitem{Misra2018}
S. Misra, M. Stern, A. Joshua, V. Umansky, and I. Bar-Joseph,
%Experimental Study of the Exciton Gas-Liquid Transition in Coupled Quantum Wells
\href{https://journals.aps.org/prl/abstract/10.1103/PhysRevLett.120.047402}{Phys. Rev. Lett. {\bf 120}, 047402 (2018)},

\bibitem{Hubert2019}
C. Hubert, Y. Baruchi, Y. Mazuz-Harpaz, K. Cohen, K. Biermann, M. Lemeshko, K. West, L. Pfeiffer, R. Rapaport, and P. Santos,
%Attractive Dipolar Coupling between Stacked Exciton Fluids
\href{https://journals.aps.org/prx/abstract/10.1103/PhysRevX.9.021026}{Phys. Rev. X {\bf 9}, 021026 (2019)}.

%nonlinearity in exciton transport
\bibitem{Ivanov2002} A. L. Ivanov,
\href{https://iopscience.iop.org/article/10.1209/epl/i2002-00144-3}{Europhys. Lett. {\bf 59}, 586–591 (2002)}.

\bibitem{Winbow2011} A. G. Winbow, J. R. Leonard, M. Remeika, Y. Y. Kuznetsova, A. A. High, A. T. Hammack, L. V. Butov, J. Wilkes, A. A. Guenther, A. L. Ivanov, M. Hanson, and A. C. Gossard,
\href{https://journals.aps.org/prl/abstract/10.1103/PhysRevLett.106.196806}{Phys. Rev. Lett. {\bf 106}, 196806 (2011)}.

\bibitem{Cohen2011} K. Cohen, R. Rapaport, and P. V. Santos,
\href{https://journals.aps.org/prl/abstract/10.1103/PhysRevLett.106.126402}{Phys. Rev. Lett. {\bf 106}, 126402 (2011)}.

\bibitem{Fedichkin2015} F. Fedichkin, P. Andreakou, B. Jouault, M. Vladimirova, T. Guillet, C. Brimont, P. Valvin, T. Bretagnon, A. Dussaigne, N. Grandjean, and P. Lefebvre, 
%Transport of dipolar excitons in (Al,Ga)N/GaN quantum wells,
\href{https://journals.aps.org/prb/abstract/10.1103/PhysRevB.91.205424}{Phys. Rev. B {\bf91}, 205424 (2015)}.

\bibitem{Dorow2016} C.J. Dorow, Y.Y. Kuznetsova, J.R. Leonard, M.K. Chu, L.V. Butov, J. Wilkes, M. Hanson, and A.C. Gossard, 
%Indirect excitons in a potential energy landscape created by a perforated electrode,
\href{https://aip.scitation.org/doi/10.1063/1.4942204}{Appl. Phys. Lett. {\bf108}, 073502 (2016)}.

\bibitem{Shahnazaryan2021} 
V. Shahnazaryan, and H. Rostami
%Nonlinear exciton drift in piezoelectric two-dimensional materials
\href{https://journals.aps.org/prb/abstract/10.1103/PhysRevB.104.085405}{Phys. Rev. B {\bf 104}, 085405 (2021)}.

\bibitem{Chiaruttini2021}
F. Chiaruttini, T. Guillet, C. Brimont, D. Scalbert, S. Cronenberger, B. Jouault, P. Lefebvre, B. Damilano, and M. Vladimirova,
%Complexity of the dipolar exciton Mott transition in GaN/(AlGa)N nanostructures
\href{https://journals.aps.org/prb/abstract/10.1103/PhysRevB.103.045308}{Phys. Rev. B {\bf 103}, 045308 (2021)}.

%IX platforms
\bibitem{Butov1999}
%Photoluminescence kinetics of indirect excitons in  GaAs/AlxGa1-xAs coupled quantum wells
L. V. Butov, A. Imamoglu, A. V. Mintsev, K. L. Campman, and A. C. Gossard,
\href{https://journals.aps.org/prb/abstract/10.1103/PhysRevB.59.1625}{Phys. Rev. B 59, 1625 (1999)}.

\bibitem{Fogler2014}
M. M. Fogler, L. V. Butov, and  K. S. Novoselov, 
%High-temperature superfluidity with indirect excitons in van der Waals heterostructures
\href{https://www.nature.com/articles/ncomms5555?origin=ppub}{Nat. Comm. {\bf 5}, 4555 (2014)}.

\bibitem{Calman2018}
E. V. Calman, M. M. Fogler, L. V. Butov, S. Hu, A. Mishchenko, and A. K. Geim, 
%Indirect excitons in van der Waals heterostructures at room temperature
\href{https://www.nature.com/articles/s41467-018-04293-7}{ Nat. Comm. {\bf 9}, 1895 (2018)}.

%dipolaritons
\bibitem{Cristofolini2012}
P. Cristofolini, G. Christmann, S. I. Tsintzos, G. Deligeorgis, G.  Konstantinidis, Z. Hatzopoulos, P. G. Savvidis, and J. J. Baumberg,
%Coupling Quantum Tunneling with Cavity Photons
\href{https://www.science.org/doi/10.1126/science.1219010}{Science {\bf 336}, 704-707 (2012)}.

\bibitem{Rosenberg2018}
I. Rosenberg, D. Liran, Y. Mazuz-Harpaz, K. West, L. Pfeiffer, and R. Rapaport,
%Strongly interacting dipolar-polaritons
\href{https://www.science.org/doi/10.1126/sciadv.aat8880}{Science Advances {\bf 4},  eaat8880 (2018)}.

\bibitem{Togan2018}
E. Togan, H.-T. Lim, S. Faelt, W. Wegscheider, and A. Imamoglu,
\href{https://journals.aps.org/prl/abstract/10.1103/PhysRevLett.121.227402}{Phys. Rev. Lett. {\bf 121}, 227402 (2018)}.


\bibitem{Mitzi1994}
D. B. Mitzi, C. A. Feild, W. T. A. Harrison, and A. M. Guloy, 
%Conducting tin halides with a layered organic-based perovskite structure
\href{https://www.nature.com/articles/369467a0}{Nature {\bf 369}, 467–469 (1994)} .

%---------------------------
\bibitem{Ahmad2015} S. Ahmad, P.K. Kanaujia, H.J. Beeson, A. Abate, F. Deschler, D. Credgington, U. Steiner, G.V. Prakash, and J.J. Baumberg, 
\href{https://pubs.acs.org/doi/10.1021/acsami.5b07026}{ACS Appl. Mater. Interfaces {\bf 7}, 25227 (2015)}.

\bibitem{Straus2018} D.B. Straus, and C.R. Kagan,
\href{https://pubs.acs.org/doi/10.1021/acs.jpclett.8b00201}{J. Phys. Chem. Lett. {\bf 9}, 1434 (2018)}.

\bibitem{Li2019} S. Li, J. Luo, J. Liu, and J. Tang, 
\href{https://pubs.acs.org/doi/abs/10.1021/acs.jpclett.8b03604}{J. Phys. Chem. Lett. {\bf 10}, 1999 (2019)}.

\bibitem{Marongiu2019} D. Marongiu, M. Saba, F. Quochi, A. Mura and G. Bongiovanni, 
%The role of excitons in 3D and 2D lead halide perovskites, 
\href{https://pubs.rsc.org/en/content/articlelanding/2019/tc/c9tc04292j}{J. Mater. Chem. C {\bf 7}, 12006 (2019)}.

\bibitem{Deng2020} S. Deng, E. Shi, L. Yuan, L. Jin, L. Dou and L. Huang, 
\href{https://www.nature.com/articles/s41467-020-14403-z}{Nature Communications {\bf 11}, 664 (2020)}.

\bibitem{Tanaka2005} K. Tanaka, T. Takahashi, T. Kondo, T. Umebayashi, K. Asai, and K. Ema,  
\href{https://journals.aps.org/prb/abstract/10.1103/PhysRevB.71.045312}{Phys. Rev. B {\bf 71}, 045312 (2005)}.

\bibitem{Yaffe2015} O. Yaffe, A. Chernikov, Z.M. Norman, Y. Zhong, A. Velauthapillai, A. van der Zande, J.S. Owen, and T.F. Heinz, 
\href{https://journals.aps.org/prb/abstract/10.1103/PhysRevB.92.045414}{Phys. Rev. B {\bf 92}, 045414 (2015)}.

\bibitem{Mauck2019} C.M. Mauck, and W.A. Tisdale,  \href{https://www.sciencedirect.com/science/article/abs/pii/S2589597419300942}{Trends Chem. {\bf 1}, 380 (2019)}.

%
\bibitem{Saparov2016} B. Saparov, and D.B. Mitzi,
\href{https://pubs.acs.org/doi/10.1021/acs.chemrev.5b00715}{Chem. Rev. {\bf 116}, 4558 (2016)}.

\bibitem{Gong2018} X. Gong, O. Voznyy, A. Jain, W. Liu, R. Sabatini, Z. Piontkowski, G. Walters, G. Bappi, S. Nokhrin, O. Bushuyev, M. Yuan, R. Comin, D. McCamant, S.O. Kelley and E.H. Sargent,
\href{https://www.nature.com/articles/s41563-018-0081-x}{Nat. Mater. {\bf 17}, 550 (2018)}.

\bibitem{Blancon2018} J.-C. Blancon, A.V. Stier, H. Tsai, W. Nie, C.C. Stoumpos, B. Traoré, L. Pedesseau, M. Kepenekian, F. Katsutani, G.T. Noe, J. Kono, S. Tretiak, S.A. Crooker, C. Katan, M. G. Kanatzidis, J. J. Crochet, J. Even and A. D. Mohite,
%Scaling law for excitons in 2D perovskite quantum wells
\href{https://www.nature.com/articles/s41467-018-04659-x}{Nat. Comm. {\bf 9}, 2254 (2018)}.


\bibitem{Quan2019} L. N. Quan, B.P. Rand, R.H. Friend, S.G. Mhaisalkar, T.-W. Lee. E.H. Sargent, 
\href{https://pubs.acs.org/doi/10.1021/acs.chemrev.9b00107}{Chem. Rev. {\bf 119}, 7444 (2019)}.

%blueshift
\bibitem{Huang2017} C. Huang, Y. Gao, S. Wang, C. Zhang, N. Yi, S. Xiao, and Q. Song, 
\href{https://www.sciencedirect.com/science/article/pii/S2211285517305955}{Nano Energy {\bf 41}, 320 (2017)}.

\bibitem{Abdelwahab2019} I. Abdelwahab, P. Dichtl, G. Grinblat, K. Leng, X. Chi, I.-H. Park, M.P. Nielsen, R.F. Oulton, K.P. Loh, and S.A. Maier,
\href{https://onlinelibrary.wiley.com/doi/abs/10.1002/adma.201902685}{Advanced Materials 31, 1902685 (2019)}.

\bibitem{Ohara2019} K. Ohara, T. Yamada, H. Tahara, T. Aharen, H. Hirori, H. Suzuura, and Y. Kanemitsu, \href{https://journals.aps.org/prb/abstract/10.1103/PhysRevB.103.L041201}{Phys. Rev. Mat. {\bf 3}, 111601(R) (2019)}.

%---------------------------
\bibitem{PanofskyBook}
W. K. H. Panofsky and M. Phillips, Classica/ Electricity and Magnetism (Addison-Wesley, Cambridge, 1961). 

\bibitem{Kumagai1989} M. Kumagai, and T. Takagahara,
%Excitonic and nonlinear-optical properties of dielectric quantum-well structures
\href{https://journals.aps.org/prb/abstract/10.1103/PhysRevB.40.12359}{Phys. Rev. B {\bf 40}, 12359 (1989)}.

\bibitem{Lang1973}
N. D. Lang, and W. Kohn,
%Theory of metal surfaces: induced surface charge and image potential,
\href{https://journals.aps.org/prb/abstract/10.1103/PhysRevB.7.3541}{Phys. Rev. B {\bf 7 }, 3541 (1973)}.

\bibitem{Muljarov1995}
E. A. Muljarov, S. G. Tikhodeev, N. A. Gippius, and T. Ishihara,
%Excitons in self-organized semiconductor/insulator superlattices: PbI-based perovskite compounds,
\href{https://journals.aps.org/prb/abstract/10.1103/PhysRevB.51.14370}{Phys. Rev. B {\bf 51}, 14370 (1995)}.



\bibitem{Ciuti1998}
C. Ciuti, V. Savona, C. Piermarocchi, A. Quattropani, and P. Schwendimann,
%Role of the exchange of carriers in elastic exciton-exciton scattering in quantum wells
\href{https://journals.aps.org/prb/abstract/10.1103/PhysRevB.58.7926}{Phys. Rev. B {\bf 58}, 7926 (1998)}.

\bibitem{Tassone1999} F. Tassone, and Y. Yamamoto, 
%Exciton-exciton scattering dynamics in a semiconductor microcavity and stimulated scattering into polaritons,
\href{https://journals.aps.org/prb/abstract/10.1103/PhysRevB.59.10830}{Phys. Rev. B \textbf{59}, 10830 (1999)}.


\bibitem{Shahnazaryan2016} V. Shahnazaryan, I. A. Shelykh, and O. Kyriienko, 
%Attractive Coulomb interaction of two-dimensional Rydberg excitons
\href{https://journals.aps.org/prb/abstract/10.1103/PhysRevB.93.245302
}{Phys. Rev. B \textbf{93}, 245302 (2016)}.

\bibitem{Shahnazaryan2017}  V. Shahnazaryan, I. Iorsh, I. A. Shelykh, and O. Kyriienko, 
%Exciton-exciton interaction in transition-metal dichalcogenide monolayers,
\href{https://journals.aps.org/prb/abstract/10.1103/PhysRevB.96.115409}{Phys. Rev. B \textbf{96}, 115409 (2017)}.

\bibitem{Hahn2005} T. Hahn,
\href{https://www.sciencedirect.com/science/article/pii/S0010465505000792}{Computer Physics Communications {\bf 168}, 78 (2005)}.



\bibitem{Schindler2008} 
C. Schindler, and R. Zimmermann
%Analysis of the exciton-exciton interaction in semiconductor quantum wells
\href{https://journals.aps.org/prb/abstract/10.1103/PhysRevB.78.045313}{Phys. Rev. B 78, 045313 (2008)}.





\end{thebibliography}
\end{document}